# Reusing optical supports using a simple software

D. Quatrini[1], R. De Angelis[2]

1: Department of Electronic Engineering, Faculty of Engineering, University of Rome "Tor Vergata"

2: Department of Physics, Faculty of Mathematical, Physical and Natural Sciences, University of Rome "Tor Vergata"

Phone: +39 339 78 99 507

davide.quatrini@students-live.uniroma2.it; rdeangelis@roma2.infn.it

Abstract

In this paper we show how it is possible to reuse optical supports (CDs, DVDs, etc.) without using chemical or physical transformation, only employing a software that can easily run on domestic computers. This software can make obsolete optical supports useful again, converting de facto WEEE (Waste electric and electronic equipment) into EEE (Electric and electronic equipment). A massive use of such a software can lead to a significant change in EEE every-day use, reducing its production to sustainable levels.

*Keywords: WEEE, EEE, Optical support, Reuse*

# 1. Introduction

WEEE (Waste electric and electronic equipment) is one of the most important problems of today disposal/recycling processes [1]. The reasons are many, but the most significant are: the massive use of highly polluting, non-biodegradable substances in EEE (electric and electronic equipment) production like plastics and metals; the volume of EEE production, which is enormous, especially for the production of every-day used EEE. So we have an enormous production of EEE that translates, after weeks, months or years, into an enormous quantity of highly-polluting, non-biodegradable WEEE.

In this paper we suggest a method for reusing optical supports like compact discs, DVDs or Blue ray discs, an every-day used EEE that forms an important part of WEEE.

The method we suggest does not involve chemical or physical transformation, neither requires the use of industrial facilities. It can be employed also in an every-day life context, because it is based solely on the use of a software, whose features are described in the following sections of the present work. The use in an every-day life context is important because we know [2] that a large part of WEEE is actually stored in private houses for a long period of time.

### 1.1 What an optical support (or also a file) is (from a binary point of view)

An optical support, like a CD or a DVD, is, despite of its physical structure, a long, very long string of binary digits, i.e. a sequence of '0' and '1' with an average "length" ranging from some MBs to a few GBs. So we can consider an optical support as a large binary "file", at least from a "binary" point of view. In the present work we will always consider the optical support as its binary content, otherwise where specified. It is important to note that any file, despite its "collocation" (hard disk, solid state memory, etc.) is, from a binary point of view, exactly the same as an optical support as described above: a variable-length string of binary digits.

## 2. Definitions and hypothesis

So if we consider an optical support and a file as homogeneous objects from a binary point of view then a system of definitions can be built. We build the system through concepts taken by arithmetic, especially from set theory.

Definition one: A is **smaller** than B if, and only if, A length in bit is smaller (<) than B length in bit (e.g. if A is 3 kb and B is 1 Mb then A is smaller than B);

Definition two: A is **a member** of B if, and only if:

- A is smaller than B, AND

- the entire binary sequence that forms A can be found within the binary sequence of B without interruptions (e.g. if A is 1111 and B is 00111100 then A is a member of B; if A is 1111 and B is 00110011 then A is not a member of B, because the binary sequence of A, i.e. 1111, is present in B sequence, but it is "broken" by the 00 interruption).

In the following sections we will always consider a file that we want to save (and we will call it 'A') and an optical support that is obsolete (WEEE) and we will call it 'B'. Besides we will assume that 'A' is a member of 'B'.

### 2.1 Describing the starting situation

Let us imagine the following starting situation: A is in our hard disk drive and it is 1 Gb long. B is a waste optical support that contains 4 Gb of obsolete data. So we have 1 Gb of occupied hard disk space and a piece of WEEE that, if not correctly recycled, will pollute the environment with plastic and metal. Now we put B in our optical support drive and run the software that we describe in the following paragraph.

## 3. Our software pseudocode

The software will act in the following way:

Step 1: identify "where" A is contained B. For doing this only the position of the starting bit and the position of the last bit is needed. Let us suppose that the binary sequence of A starts from the bit N° 2,534,988,331 of B and ends on the bit N° 3,534,988,331 (as we said before A is formed by 1 Gb long string of bits and, because it is a member of B, A starts from a specific bit of B and ends on another specific bit of B);

Step 2: produce a file (we call it 'C') containing only the ordinal number of the starting bit and the ordinal number of the ending bit. Because of B length a file of only 64 bits is needed for doing this (32 bits for the starting ordinal number and 32 bits for the ending ordinal number);

Step 3: delete A from our hard disk. Now we have 1 Gb of free hard disk space.

Please note that the A file is not lost. Our software can easily rebuild it in the following way:

Step a): load C;

Step b): use information contained in C for extracting A from B.

### 3.1 Results

The application of the described software leads to the following results:

1) obsolete optical supports are reused as "bit depots" for building useful files, so they are no more obsolete (WEEE reverts to EEE);

2) hard disks and non-volatile memories in general are freed from files, replaced by much more smaller "index" files (as the C file cited above);

3) because of 1) and 2) EEE production is reduced, so its disposal and, consequently, environmental pollution.

## 4. Problems, solutions and features

The main problem of such a software is probability. It is very improbable to find a 1 Gb A file that is a member of a 4 Gb B file. Probably our entire domestic collection of obsolete optical supports it is not sufficient for finding a B that successfully contains an A (unless in the case of an A made of a very short string of bits).

This problem can be easily overcome subdividing A into subfiles (A1, A2, ..., An). The sum of the obtained C1, C2, ..., Cn files will be smaller than the entire A file, making the use of the described software convenient in most cases.

Following this path we can also say that a B optical support can easily "contain" more than one 'A' file. Using different "index" files (i.e. 'C' files) we can "encode" different 'A' files (not necessarily logically connected or subfiles of a larger file) into a single 'B' optical support. Our software will use different 'C' files for rebuilding the different 'A' files, but it will use always the same 'B' optical support, de facto multiplying the benefits described above in section 5. Besides the same approach can be used for reusing other kinds of memories (not only optical supports) leading to a general more efficient way of EEE use.